\documentclass[aps,prc,twocolumn,floatfix,superscriptaddress,showpacs,showkeys]{revtex4-1}

\usepackage{hyperref}
\usepackage{dcolumn}
\usepackage{graphicx}

\usepackage{natbib}
\usepackage{gensymb}

\usepackage{amsmath}
\usepackage{tikz}
\usepackage{verbatim}

\usepackage{graphicx}
\usepackage{multirow}
\newcolumntype{+}{D{+}{\,\pm\,}{3,3}}

\newcommand{\ganil}{{ GANIL, CEA/DSM-CNRS/IN2P3, BP55027, F14076 Caen
    Cedex 5, France}}%

\newcommand{\lpc}{{LPC Caen, ENSICAEN, Universit\'e de Caen,
    CNRS/IN2P3, F14050 CAEN Cedex, France}}%

\newcommand{\ipno}{{ IPN Orsay, IN2P3-CNRS, Universit\'e Paris Sud,
    F91406 Orsay, France}}%

\newcommand{\saclay}{{ CEA-Saclay, DSM/IRFU SPhN, F91191
    Gif-sur-Yvette Cedex, France}}%

\newcommand{\dubna}{{ Flerov Laboratory of Nuclear Reactions, JINR,
    Dubna, RU141980 Russia}}%

\newcommand{\warsaw}{{National Centre for Nuclear Research,  ul.\ Andrzeja So\l tana 7, 05-400 Otwock, Poland }}%

\newcommand{\HIL}{Heavy Ion Laboratory, University of Warsaw, ul.\ Pasteura 5A, PL-02-093 Warsaw, Poland.}

\begin{document}

\title{Structure of unbound neutron-rich $^9$He studied using single-neutron transfer}

\date{\today}

\author{T.~Al~Kalanee}
\affiliation{\lpc}
\affiliation{\ganil}

\author{J.~Gibelin}\email[Electronic address:]{gibelin@lpccaen.in2p3.fr}
\affiliation{\lpc}
\affiliation{\ganil}
\author{P.~Roussel-Chomaz}
\affiliation{\ganil}
\author{N.~Keeley}
\affiliation{\warsaw}
\author{D.~Beaumel}
\affiliation{\ipno}
\author{Y.~Blumenfeld}
\affiliation{\ipno}
\author{B.~Fern\'andez-Dom\'inguez}
\affiliation{\ganil}
\author{C.~Force}
\affiliation{\ganil}
\author{L.~Gaudefroy}
\affiliation{\ganil}
\author{A.~Gillibert}
\affiliation{\saclay}
\author{J.~Guillot}
\affiliation{\ipno}
\author{H.~Iwasaki}
\affiliation{\ipno}
\author{S.~Krupko}
\affiliation{\dubna}
\author{V.~Lapoux}
\affiliation{\saclay}
\author{W.~Mittig}
\affiliation{\ganil}
\author{X.~Mougeot}
\affiliation{\saclay}
\author{L.~Nalpas}
\affiliation{\saclay}
\author{E.~Pollacco}
\affiliation{\saclay}
\author{K.~Rusek}
\affiliation{\warsaw}
\affiliation{\HIL}
\author{T.~Roger}
\affiliation{\ganil}
\author{H.~Savajols}
\affiliation{\ganil}
\author{N.~de S\'er\'eville}
\affiliation{\ipno}
\author{S.~Sidorchuk}
\affiliation{\dubna}
\author{D.~Suzuki}
\affiliation{\ipno}
\author{I.~Strojek}
\affiliation{\warsaw}
\author{N.~A.~Orr}
\affiliation{\lpc}

\begin{abstract}

The $^8$He(d,p) reaction was studied in inverse kinematics at
15.4$A$~MeV using the MUST2 Si-CsI array in order to shed light on the
level structure of $^9$He. The well-known $^{16}$O(d,p)$^{17}$O
reaction, performed here in reverse kinematics, was used as a test to
validate the experimental methods. The $^9$He missing mass spectrum
was deduced from the kinetic energies and emission angles of the
recoiling protons.  Several structures were observed above the
neutron-emission threshold and the angular distributions were used to
deduce the multi-polarity of the transitions. This work confirms that
the ground state of $^9$He is located very close to the neutron
threshold of $^8$He and supports the occurrence of parity inversion in
$^9$He.

\end{abstract}

\maketitle

\section{Introduction}

Neutron-rich $N=7$ isotones are of particular interest because of
their level sequence, which differs from that predicted by the shell
model for nuclei near stability \cite{Talmi:n7}.
The standard shell model generally predicts a $J^\pi = 1/2^-$ ground state
(G.S.) for $N=7$ nuclei. This is true for $^{15}$O and $^{13}$C, but
$^{11}$Be presents parity inversion with a $1/2^+(\nu 2s_{1/2})$
ground state~\cite{deu68}. This parity inversion was predicted for the
first time by Talmi and Unna in 1960~\cite{Talmi:n7}: they showed that
parity inversion for $^{11}$Be can be predicted from
linear extrapolation of the $p_{1/2}$-$s_{1/2}$ energy in
$^{13}\text{C}$ ($3.09$~MeV) and the corresponding difference between
the center of mass associated states in $^{12}\mathrm{B}$
($1.44$~MeV). Hence the $s_{1/2}$ state was predicted to be the G.S. of
$^{11}\text{Be}$ at $0.21$~MeV below the $p_{1/2}$ level.  Recent
results for $^{10}$Li~\cite{sim04,jep06,alfalou:n7:conf,aks08} also
confirm the observation of a virtual $s$ state close to the neutron
emission threshold and the presence of a resonance around $0.5$~MeV.

The first results for the unbound $^9$He nucleus were obtained by
Seth~\textit{et al.} in 1987 \textit{via} the double-charge exchange
reaction $^{9}\text{Be}(\pi^-,\pi^+)^9\text{He}$
\cite{Seth:1987kq}. The lowest energy state observed was considered to
be the ground state at $1.13(10)$~MeV above the neutron threshold with
a width of $\Gamma = 0.42(0.1)$~MeV and a $1p_{1/2}$ configuration.
Two excited states were observed: the first excited state was
identified as a $2s_{1/2}$ state at $2.33(0.1)$~MeV with $\Gamma =
0.42(0.1)$ and the second as a $5/2^+$ or $3/2^-$ state at
$4.93(0.1)$~MeV ($\Gamma = 0.5(0.1)$~MeV). There was also a possible
state at $8.13$~MeV with $\Gamma = 0.55(0.1)$~MeV.

The $^{9}\text{Be}(^{13}\text{C},^{13}\text{O})$ reaction was studied
by von~Oertzen~\textit{et~al.}~\cite{oer95} and
Bohlen~\textit{et~al.}~\cite{boh88}.  Despite low statistics, a state
at $1.13$~MeV above the neutron threshold and another state at
$4.93$~MeV were observed.  The same authors \cite{oer95,Bohlen:1999qq}
also investigated the
$^{9}\text{Be}(^{14}\text{C},^{14}\text{O})^9\text{He}$ reaction. A
$J^\pi = 1/2^-$ state was proposed for the ground state at $1.27$~MeV
above the neutron threshold with $\Gamma = 0.1(6)$~MeV. Three excited
states were found at $2.37(10)$ (with $\Gamma = 0.7(2)$~MeV),
$4.3(10)$ and $5.25(10)$~MeV, respectively.  Note that recently the
heavy-ion double-charge exchange reaction
$^{9}\text{Be}\left(^{18}\text{O},^{18}\text{Ne}\right)\protect{^{9}}\text{He}$
\cite{matsubara:he9:proc} was investigated but the results do not show
any structure in $^9$He.

In these three studies, the state identified as the ground state was
assigned a $J^\pi={1/2^-}$ spin-parity, leading to the conclusion that
there is no parity inversion in $^9$He, thus breaking the systematics
started with $^{11}\text{Be}$ and $^{10}\text{Li}$. These results were
consistent with theoretical studies at the time \cite{pop85,ogl95},
however they are in contradiction with more recent calculations
\cite{sag93,kit93,pop93,Chen:2001qy,Otsuka:PRL87}.

More recently Barker \cite{bar04} showed that the small width of the
$1/2^-$ level ($\Gamma=0.42$~MeV \cite{Seth:1987kq} and
$\Gamma=0.1$~MeV \cite{oer95}) is inconsistent with a single-particle
state. According to Barker's calculations, the $1/2^-$ single-particle
width for $^8\text{He}+\text{n}$ should be about $1$~MeV.\\

The two-proton knock-out reaction from $^{11}\text{Be}$ at 28$A$~MeV
studied by Chen~\textit{et~al.}  \cite{Chen:2001qy} was the first
experiment to identify a state at a lower energy than the earlier
experiments. This ``new'' ground state at around $0.2$~MeV above the
$^8\text{He}+\text{n}$ threshold was assigned a $2s_{1/2}$
configuration, indicating for the first time parity inversion in the
$^9$He nucleus. The scattering length $a_s$ found by Chen was $a_s
\leq-10$~fm, corresponding to a virtual state of energy $E_r \lesssim
0.2$~MeV. These results were consistent with shell model calculations
undertaken by Warburton and Brown \cite{war92}.

Later, the C$(^{11}\text{Be},^8\text{He}+\text{n})$X reaction at
35$A$~MeV was studied by
Al~Falou~\emph{et~al.}~\cite{alfalou:phd,alfalou:n7:conf}. The
neutron-$^8$He relative energy spectrum could be explained by a
virtual $s$ state of scattering length $-3 \lesssim a_s \lesssim
0$~fm, consistent with no or at most a very weak final state
interaction.  Al~Falou~\emph{et~al.} also studied the
C$(^{14}\mathrm{B},^8\text{He}+\text{n})$X reaction at 35$A$~MeV the
results of which were consistent with a very weakly interacting
$s$-state and a resonance at $E_r \approx 1.3$~MeV with a width of
$\Gamma \approx 1$~MeV.

The most recent results on $^9\text{He}$ using this type of reaction
were published by Johansson
\textit{et~al.}\ \cite{Johansson:2010fk}. They used the
$^{1}\text{H}(^{11}\mathrm{Li},^{8}\mathrm{He}+n)$ knock-out reaction
at 280$A$~MeV.  Their work shows dominant $s$-wave scattering at low
energy with $a_s = -3.17(66)$~fm in addition to two resonances at
$1.33(8)$~MeV ($\Gamma=0.1(6)$~MeV) and $2.42(10)$~MeV ($\Gamma =
0.7(20)$~MeV) above the neutron threshold.  Given the reaction used
and the very small scattering length, the low energy structure was
attributed to a threshold effect rather that a true state.

A study of the isobaric analog states of $^9$He in $^9$Li was
performed by Rogachev~\textit{et al.}~\cite{Rogachev:2003lq}
\textit{via} the elastic scattering of $^8$He on protons at around
7$A$~MeV. The experimental setup did not allow the lowest energy states
to be observed but three states were seen above threshold: a $1/2^-$
or $3/2^-$ state at $1.1$~MeV with a width of $\Gamma < 0.1$~MeV, a
second $3/2^-$ (or $1/2^-$) state at $2.2$~MeV ($\Gamma =
1.1(0.4)$~MeV), and a $(5/2^+,3/2^+)$ state at $4.0$~MeV with $\Gamma
= 0.24(0.1)$~MeV.

Finally, the structure of $^9$He has been studied using transfer
reactions. Fortier \textit{et~al.}~\cite{Fortier2007} employed the
d($^8$He,p)$^9$He reaction at a beam energy of $15.3A$~MeV.  The use
of eight MUST~\cite{yorick:must1} telescopes at backward angles
enabled laboratory angles $\left(\theta_{lab}\right)$ from $110^\circ$
to $170^\circ$ to be covered.  This study found three states at
$\approx 0$~MeV, $1.3$~MeV and $2.3$~MeV and two other possible states
at higher energies~\cite{fortier:9He}.  The angular distributions
measured for the first two states suggest an inversion between the
$1/2^+$ and $1/2^-$ levels, but the very limited statistics obtained
(especially for the peak around the neutron threshold) made it
difficult to draw definite conclusions.

The same reaction was studied by Golovkov~\textit{et
  al.}~\cite{Golovkov:2007fu}, using a $^8$He beam at the higher
energy of $25A$~MeV. The presence of a virtual state with a
scattering length $a_s > -20$~fm was inferred from the large
forward-backward asymmetry of the spectra.  Two excited states were
observed: a $1/2^-$ state at $2$~MeV with $\Gamma \sim 2$~MeV and a
$5/2^+$ state at $\geq 4.2$~MeV with $\Gamma > 0.5$~MeV.  It is worth
noting that the width of the first excited state in this work is much
larger than in the majority of the previous experiments.

The present experiment used the same reaction and an almost identical
energy to Fortier~\textit{et al.}~\cite{Fortier2007} ---
$\text{d}\left(^{8}\text{He},^{9}\text{He}\right)\text{p}$ at
15.4$A$~MeV --- and benefited from an increased of the $^{8}\text{He}$
beam intensity and an improved angular coverage possible with the new
MUST2 array \cite{loly:enam05}.

\begin{figure}[!ht]
\includegraphics[width=\columnwidth]{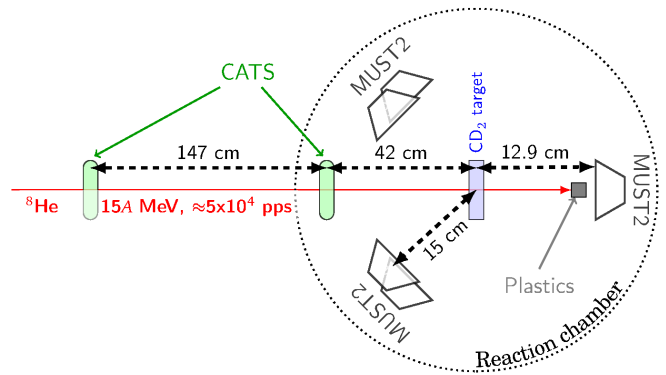}
\caption{(color online). Schematic diagram of the experimental
  setup\label{fig:setup}. Trapezium shapes represent MUST2 modules.}
\end{figure}

\section{Experiment\label{sec:setup}}

The experiment was performed at the GANIL facility and was part of the
first MUST2 campaign \cite{daisuke:12o:prl,Mougeot2012}. A secondary
$^8$He beam at 15.4$A$~MeV was produced by the SPIRAL1 ISOL
facility~\cite{Villari:SPIRAL1} \textit{via} the fragmentation of a
75$A$~MeV $^{13}$C beam in a thick carbon target.  After
re-acceleration and purification the beam was delivered to a
deuterium-enriched polyethylene targets (CD$_2$)$_n$ of
320~$\mu$g/cm$^2$ or 550~$\mu$g/cm$^2$ thickness. The horizontal and
vertical size of both targets was 4~cm and 3~cm respectively. The
experimental setup is shown in Fig.~\ref{fig:setup}.  For more details
see Ref.~\onlinecite{Tarek:PhD,daisuke:phd,daisuke:phd:epja,mougeot:phd}.

The beam spot on the target and the incident angles of incoming
particles were monitored event-by-event using two sets of multi-wire
low pressure chambers, CATS \cite{ott99}. The typical size of the
$^8$He beam was 3.3~mm (FWHM) and the range of incoming angles was
26~mrad (FWHM).

The energies and angles of the recoiling protons were measured by an
array of four MUST2 telescopes \cite{loly:enam05} located upstream of
the target. Each telescope, with an active area of $10\times
10$~cm$^2$, consisted of a 300-$\mu$m~thick double-sided Si strip
detector (DSSSD) and a 4-cm~thick 16-fold CsI calorimeter, which
provided energy-loss ($\Delta$E) and residual-energy (E) measurements,
respectively. The DSSSDs were divided into 128 strips in both the $x$
and $y$ directions, thus providing position information. The emission
angle of the recoiling particles was obtained by combining this
information with the angle and position of the incoming $^8$He on the
target.

The acceptance of the array was estimated using a Monte-Carlo
simulation which took into account the detector geometry and the beam
profile.  For the proton at backward angle, the setup covered
laboratory (center-of-mass) angles between $120\degree-170\degree$
($\sim~2.7\degree-21.4\degree$). The acceptance has a maximum value of
$\sim 80\%$ at $\theta_\mathrm{lab} = 135\degree-160\degree$
($\theta_\mathrm{c.m.} \sim~6\degree-16\degree$), while it gradually
decreases toward smaller or larger angles.  The total kinetic energy
was obtained from the proton energy information, to which a correction
was applied based on the calculated energy loss in the target. This
correction depends on both energy and angle and was typically of the
order of 50~keV.

The beam particles and forward emitted $^8$He from the in-flight decay
of $^9$He were detected by a $20\times20$~mm$^2$, 1~mm thick, NE102
plastic scintillator located 11~cm downstream of the target and
covering $\theta_\mathrm{lab} = 0\degree$ to 5.6\degree (representing
97\% of the reconstructed events). Larger angles up to
6.5\degree\ were covered by a fifth MUST2 telescope located 19~mm
behind the plastic scintillator.

\section{Analysis}

As noted above, the reaction position on the target was reconstructed
from the position measurements made using the two CATS detectors on an
event by event basis. Only events where both CATS had fired were
selected for analysis in order to reconstruct the trajectory of the
beam particle and to ensure that the beam hit the target.

The ranges of the protons of interest (emitted from the transfer
reactions populating either $^{17}$O or $^{9}$He) being such that all
were expected to stop in the first 300~$\mu$m stage of the telescope,
we rejected all events where any CsI directly behind the measured
impact position in a DSSSD had fired.  Proton identification was
performed using the energy-time of flight method~\cite{Tarek:PhD}.

The presence of the plastic scintillator located downstream of the
target allowed us to perform coincidences with all beam-like particles
and potentially outgoing $^8$He (no isotope separation was however
possible). The $^8$He beam ions that stopped in the plastic
scintillator emitted $\beta$ particles through their decay
($T_{1/2}\sim120$~ms) and produced reaction induced protons that could
be seen in the backward telescopes. The $\beta$ particles being of low
energy and uncorrelated to the beam they could be easily identified
and eliminated at the cost of losing very low energy protons below
600~keV, close to the detection threshold of MUST2.  The second
contamination (protons) was in coincidence with the beam but as
plastic scintillator was located 11~cm after the $CD_2$ target,
these particles came at least 5~ns after the protons of interest. The
MUST2 time resolution of 500~ps was then sufficient to disentangle
them, provided they did not punch through the first layer of the
telescope.

Additional conditions were employed: we selected events with
multiplicity one for all backward telescopes; we rejected events where
the energies collected from the two sides of the DSSSD were not equal
(within the resolution) and, finally, the proton energies were
restricted to kinematically reasonable ones, within the experimental
resolution.

The scattering angle was deduced from the proton hit position on
MUST2, the reaction point on target and the angle of incidence of the
beam.  The excitation energy, $E_4^*$, was calculated using:

\begin{widetext}
\begin{equation}
  E_4^* = \sqrt{(T_1+m_1+m_2-T_3-m_3)^2 - P_1^2 - P_3^2 + 2\cdot P_1
    \cdot P_3 \cdot cos(\theta) } - m_4 \label{eq:excitation}
\end{equation}
\end{widetext}

where the indices $i=1,2,3,4$ stand respectively for the beam,
deuteron, proton and nucleus of interest ($^9$He or $^{17}$O), and
$P_i$ are the momenta, $T_i$ the kinetic energies and $m_i$ the rest
masses. The masses were taken from Ref.~\cite{nndc:A16A17}
($A=16\ \mathrm{and}\ 17$) and Ref.~\cite{Tilley:2004qa} ($^8$He). The
mass of the $^9$He was defined as the sum of the rest masses of $^8$He
and a free neutron (see below).

\begin{figure}[!t]
\includegraphics[width=\columnwidth]{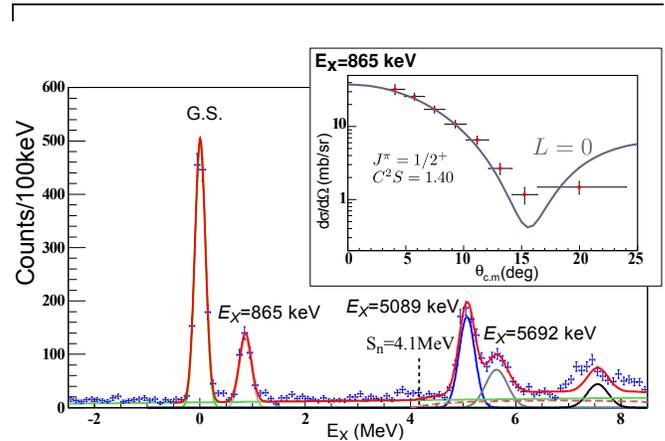}
\caption{(color online). \label{fig:o17:ex}Experimental missing mass
  spectrum for the
  $^{16}\text{O}\left(d,p\right)\protect{^{17}}\text{O}$ reaction in
  inverse kinematics. Dashed line (brown): 3-body phase space. Thin
  solid line (green): background from reactions of the beam with the
  carbon of the CD$_2$ target. Inset: sample angular distribution for
  the first excited state, compared to a DWBA calculation.}
\end{figure}

\begin{table*}[!t]
  \begin{ruledtabular}
\begin{tabular}{c+c+c}
  \multirow{2}{*}{$J^{\pi}$} & \multicolumn{1}{c}{$E_x$(keV)} & $C^2S$  & \multicolumn{2}{c}{This work} \\%
 \cline{4-5} \rule{0pt}{3ex} &
  \multicolumn{1}{c}{(Adopted values \cite{nndc:A16A17})} &
  \multicolumn{1}{c}{(Ref.~\cite{Cooper:1974xq,Darden:1973rr})} &
  \multicolumn{1}{c}{$E_x$}& \multicolumn{1}{c}{$C^2S$} \\%
 \hline%
 $5/2^+$ & \multicolumn{1}{l}{$~~~~~~~~~0$} & $1.07-0.84$ &   5+2  & 0.7  \\%
 $1/2^+$ &                      870.73+0.10 & $1.14-0.91$ & 865+9  & 1.4  \\%
 $3/2^+$ &                       5084.8+0.9 & $1.2$       & 5089+1 & 0.8  \\%
 $7/2^-$ &                     5697.26+0.33 & $0.15$      & 5692+7 & 0.13 \\%
\end{tabular}
  \end{ruledtabular}
\caption{\label{tab:17O}Comparison between the results for
  $^{16}\text{O}\left(d,p\right)\protect{^{17}}\text{O}$ in inverse
  kinematics and the adopted excitation energies ($E_x$) and
  previously published spectroscopic factors ($C^2S$) for the observed
  states. The uncertainties on the $C^2S$ for this work here are
  estimated to be of the order of $\pm 20\%$ (see text).}
\end{table*}

In order to test both our understanding of the setup and our analysis
procedures, a test measurement of the
$^{16}\text{O}\left(d,p\right)\protect{^{17}}\text{O}$ reaction at
15.5$A$~MeV was made in inverse kinematics. Note that in this case, the
MUST2 module and the plastic detector at 0$\degree$ were removed and
no recoil identification was possible.  Figure~\ref{fig:o17:ex}
presents the missing mass spectrum of $^{17}$O obtained with the
550~$\mu$g/cm$^2$ target. Two states around 0 and 0.9~MeV are clearly
separated, and two other states around 5.5~MeV can clearly be
distinguished. Although other structures are present at higher
energies we focus on these four states for which results from transfer
the same reactions are available in the literature
\cite{Darden:1973rr,Cooper:1974xq}. To refine our analysis we took
into account two physical backgrounds: a 3-body phase space simulating
the deuteron break-up and a background due to reactions of the
$^{16}$O beam with the carbon present in the CD$_2$ target.

Using Gaussian distributions for the states below the neutron
threshold $S_n = 4.1$~MeV and ``Voigt profiles'' \cite{[A convolution
    of a Lorentzian and a Gaussian distribution{,} see for
    example:\ ][]Voigt:calc} for the states above, we obtained the
energies listed in Tab.~\ref{tab:17O}. We compared these results with
tabulated and previously published values, in particular the work of
Darden~\textit{et~al.}~\cite{Darden:1973rr} and
Cooper~\textit{et~al.}~\cite{Cooper:1974xq} (focusing on the latter as
having the closest comparable conditions to our experiment
\textit{i.e.}\ 18$A$~MeV incident energy deuteron beam). We conclude
from the energies listed in this table that our setup, calibration and
analysis procedure reproduce with an accuracy better or equal to 5~keV
the energies of the ground state and the first three excited
single-particle states of $^{17}$O. In addition, taking into account
the experimental resolution deduced from simulations, the $\Gamma =
70$~keV width of the 5084~keV unbound state is reproduced.

For each state, we determined the integral above the background for a
range of c.m. angles.  Taking into account the beam exposure and
correcting for the acceptance and dead time, differential angular
distributions were constructed.  Standard finite-range Distorted Wave
Born Approximation (DWBA) calculations were carried out using the code
{\sc Fresco}~\cite{Thompson:FRESCO}.  These employed the same global
deuteron and neutron optical model parameters for the distorting
potentials in the entrance and exit channels, respectively, as used in
the d($^8$He,p) calculations described later. The binding potentials
for the $\langle d|n+p\rangle$ and
$\langle^{17}\mathrm{O}|^{16}\mathrm{O}+n\rangle$ overlaps also
employed the same parameters as for the d($^8$He,p) calculations, the
$^{16}\mathrm{O} + n$ values being similar to those of
refs.~\cite{Cooper:1974xq} and \cite{Darden:1973rr}.

The four angular distributions obtained in the test measurement were
well reproduced and the resulting angular distribution for the first
excited state is shown, as an example, in the insert of
Figure~\ref{fig:o17:ex}. This demonstrates the validity of our
experimental approach. Note that a systematic $10\%$ error was
assigned to the cross sections to take into account the effect of the
uncertainties in: target thickness, detector efficiencies, and solid
angles. 

Finally, the corresponding spectroscopic factors $C^2S$ were deduced
by the normalization of the DWBA calculations to the measured angular
distributions. The error on the normalization due to statical and
systematic errors are $\sim 10\%$, and the uncertainties arising from
the choice of potential in the DWBA calculations are estimated to be
$\sim 20\%$ (\cite{Lee:C2S}). Table~\ref{tab:17O} lists the $C^2S$,
which are in reasonable agreement with those taken from the
literature.

\section{Results}

 \begin{figure}[!ht]
 \centering \includegraphics[width=\columnwidth]{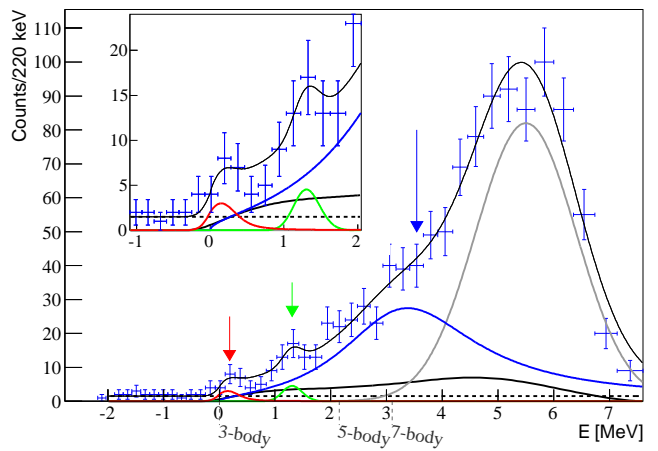}
 \caption{(color online). Experimental missing mass spectrum for for
   the $\left(p,^{8}\text{He}\right)p$ reaction which is described
   with three states: ground state (red), first excited (green) and
   second excited (blue) states. The solid gray line models the
   acceptance cutoff. The solid black line denotes the sum of the 3, 5
   and 7-body phase space contributions, whose respective breakup
   energies are noted. The dotted line indicates the physical
   background due to reactions of the beam with the plastic
   scintillator. The thin solid line is the sum of all
   contributions. The region around the threshold is shown in the
   inset.}
 \label{fig:he9:ex}
 \end{figure} 

The analysis procedure for the
$d\left(^{8}\text{He},\protect{^{9}}\text{He}\right)p$ measurement
were identical to the test experiment.  The missing mass spectrum is
presented in Fig.~\ref{fig:he9:ex}. Since the experimental resolutions
obtained with the 320 and 550~$\mu$g/cm$^2$ targets were similar, we
present here the sum of the spectra obtained with both targets. Note
that here the rest mass $m_4$ in Eq.~\ref{eq:excitation} is defined as
the sum of the $^8$He and free neutron rest masses. The calculated
missing mass energy (denoted here as $E_r$) is thus defined from the
neutron threshold of $^9$He.

Two peaks can clearly be seen: one approximately 200~keV above
threshold which we identified as the ground state (G.S.)  and another
around 1.5~MeV. We also observe a shoulder around 3~MeV. Given that
the broad structure around 6~MeV is related to the proton energy
cut-off and therefore not a real state, we concluded that the shoulder
around 3~MeV is due to a second excited state. The presence and the
position of the two excited states are compatible with several
previous reports
\cite{Seth:1987kq,oer95,Rogachev:2003lq,fortier:9He,Fortier2007}.

Using these energies as a first estimate of the resonance energies of
the states, a fit was performed employing ``Voigt profiles''
\cite{Voigt:calc}. The Lorentzian widths $\Gamma$ are energy dependent
$\Gamma = \Gamma_0\sqrt{\frac{E}{E_R}}$ \cite{rmatrix}. The Gaussian
component takes into account the experimental energy resolution. The
widths were deduced from the results for the
$^{16}\text{O}\left(d,p\right)$ test measurement at the corresponding
proton energy. Only the last structure around 6~MeV was considered as
a simple Gaussian function. Physical backgrounds associated with 3, 5
and 7-body phase spaces corresponding to $^8$He+d $\to$ $^8$He+p+n,
$^6$He+p+3n and $^4$He+p+5n were taken into account. Finally, a linear
background arising from reactions of the beam with the carbon of the
target and reactions in the plastic scintillator beam stopper was
added. The results are listed in Tab.~\ref{tab:9He}.

\begin{table*}
  \begin{ruledtabular}
  \begin{tabular}{++cccc}
    \multicolumn{1}{c}{\multirow{2}{*}{$E_r$(keV)}}  &   
    \multicolumn{1}{c}{\multirow{2}{*}{$\Gamma$(keV)}}  &   \multicolumn{4}{c}{$C^2S$} \\
    \cline{3-6}
    \multicolumn{2}{l}{} & $p_{1/2}$  &$p_{3/2}$  &$d_{3/2}$  & $d_{5/2}$  \\
    \hline
      180+85  &  180+160  &    &    &    &   \\     
     1235+115 &  \multicolumn{1}{c}{$130\,^{+~170}_{-~130}$}  & $0.02-0.05$ & $0.01-0.03$ & $0.006-0.01$ & $0.005-0.007$ \\
     3420+780 & 2900+390  &  &  & $0.03-0.04$ & $0.02-0.03$\\
  \end{tabular}
\end{ruledtabular}

  \caption{\label{tab:9He}Position and width of $^9$He states obtained
    in this work. Spectroscopic factors were deduced using DWBA
    calculations (see text), the uncertainties originate from the
    different Woods-Saxon n + $^8$He binding potentials used in the
    calculations.}
\end{table*}

We will discuss these results in detail in the next section, although
we note here that the ground state of $^9$He is found at
$180\pm85$~keV above the neutron threshold. This is compatible with
the other $\left(d,p\right)$ reaction
\cite{Golovkov:2007fu,fortier:9He} studies.  Both the position and the
rather small width of the first excited state are also compatible with
several previous experiments
\cite{Bohlen:1999qq,oer95,Rogachev:2003lq,fortier:9He,Johansson:2010fk}.
The second excited state is slightly higher in energy than the average
of previous experimental observations
\cite{Seth:1987kq,oer95,Rogachev:2003lq,fortier:9He,Golovkov:2007fu,Johansson:2010fk}.
This could be due to the uncertainty in the shape of the structure
located at 6~MeV, and a deviation of up to several hundred keV might
be possible. The error on its position has been estimated by assuming
different widths for the 6~MeV structure.

\begin{figure}[!ht]
\includegraphics[angle=-90,clip,width=\columnwidth]{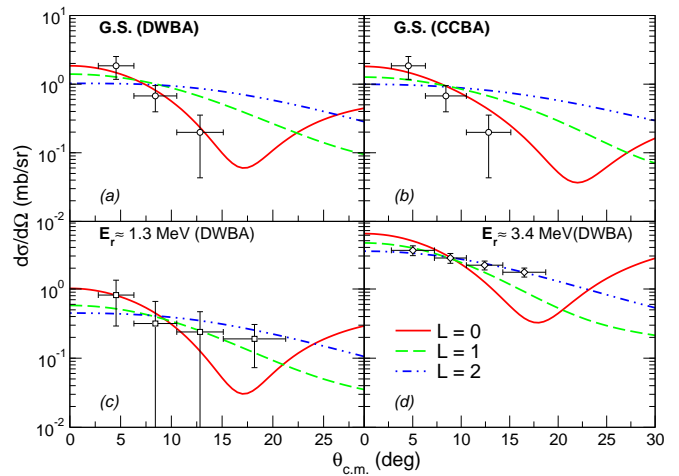}
\caption{(color online). Angular distributions for the ground state
  (a) and the two first excited states of $^9$He (c and d) compared to
  $L=0,1,2$ (respectively red, green and blue) DWBA
  calculations. Panel (b): angular distribution of the G.S. compared
  to CCBA calculations. \label{fig:ad:he9}}
\end{figure}

The experimental angular distributions for these three states are
presented in Fig.~\ref{fig:ad:he9}.  Error bars take into account
uncertainties due to the subtraction of the different backgrounds.
These angular distributions are compared with the results of full
finite range DWBA calculations, similar to those carried out for the
d($^{16}$O,p)$^{17}$O reaction. The normalization for each energy and
transferred angular momentum $L$ was obtained with a log-likelihood
fit.

The entrance channel d + $^8$He optical model potential was calculated
using the global parameters of Daehnick {\em et al.\/}~\cite{Dae80}
and the exit channel p + $^9$He potentials employed the systematics of
Koning and Delaroche~\cite{Koning2003}. The deuteron internal wave
function, including the small $D$-state component, was calculated
using the Reid soft-core interaction \cite{Rei68} as the
neutron-proton binding potential. We used the weak binding energy
approximation (WBEA) where the $^9$He internal wave functions were
calculated by binding the neutron to the $^8$He core with a standard
Woods-Saxon potential with reduced radius $r_0 = 1.25$~fm, and
diffusivity $a_0 = 0.65$~fm, the well depths being adjusted to give a
binding energy of 0.0001~MeV in all cases.  Note that test
calculations were performed with different sets of $(r_0, a_0)$
values with ranges of $1.25-1.50$ and $0.65-0.75$, respectively
without noticeable effect on our conclusions.

We chose to employ the WBEA to calculate the n + $^8$He overlaps for
two reasons: firstly, when the unbound neutron is in a relative
$s$-state with respect to the $^8$He core this results in a virtual
state rather than a conventional resonance, due to the absence of
either a Coulomb or a centrifugal barrier in the ``binding''
potential, thus rendering a more sophisticated modeling of the form
factor for such states problematic. Secondly, while states with $L >
0$ may be modeled in {\sc Fresco} as true resonances with finite
widths, in practice it is often difficult to achieve consistent
results using this procedure. We therefore chose to use the WBEA to
calculate all the n+$^8$He overlaps for the sake of consistency.
  
The procedure adopted was to perform calculations assuming angular
momentum $L$ = 0, 1 and 2 for the neutron relative to the $^8$He core
for all three states, and to compare the resulting angular
distributions to the experimental points to deduce the best-fit values
of $L$, thus providing clues as to the spin-parities of the respective
states in $^9$He, as well as spectroscopic factors. All calculations
included the full complex remnant term and thus yielded identical
results for either post- or prior-form DWBA.

Since the incident $^8$He energy is relatively high, the influence of
deuteron breakup effects could be important. To test this we performed
a coupled-channels Born approximation (CCBA) calculation for stripping
to the ground state of $^9$He. The CCBA calculation was similar in all
respects to the DWBA calculations with the exception that the entrance
channel optical potential was replaced by a continuum discretized
coupled channels (CDCC) calculation similar to that described in
Ref.~\cite{Keeley2004}. The necessary diagonal and transition
potentials were calculated using Watanabe-type folding based on the
global nucleon optical potential of Ref.\ \cite{Koning2003} and the
deuteron internal wave function of Ref.~\cite{Rei68}. As
Fig.~\ref{fig:ad:he9}~(b) shows, the shapes of the $L=0, 1, 2$ angular
distributions are almost identical to those for the corresponding DWBA
calculations, suggesting that the influence of deuteron breakup on the
shape of the angular distribution is small in this case, justifying
our use of the DWBA to infer spins and parities.

\begin{figure*}[!ht]
\includegraphics[clip]{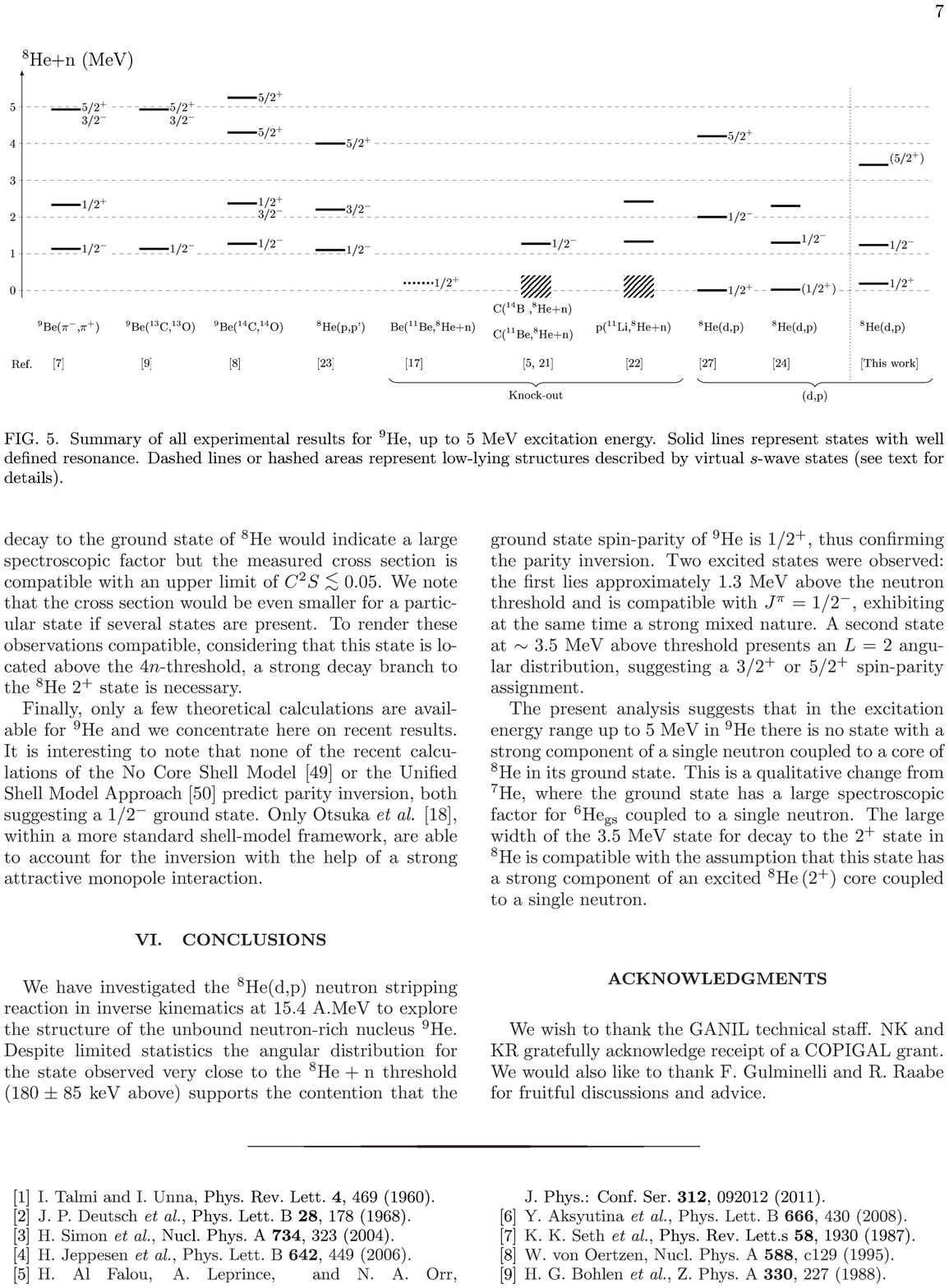}
\caption{Summary of all experimental results for $^9$He, up to 5~MeV
  excitation energy. Solid lines represent states with well defined
  resonance. Dashed lines or hashed areas represent low-lying
  structures described by virtual $s$-wave states (see text for
  details).\label{fig:he9:sum}}
\end{figure*}

\section{Discussion}

We present the $^9$He states obtained in the present work together
with all published results in Fig.~\ref{fig:he9:sum}.  This confirms
the presence of a state in $^9$He very close ($\sim$~200~keV) to the
neutron emission threshold --- previously observed in (d,p) reactions
\cite{Fortier2007,Golovkov:2007fu} --- that we have identified as the
ground state. Different theoretical angular distributions for this
state assuming different transferred angular momenta ($L = 0,1\text{
  or }2$) calculated in both the DWBA and the CCBA formalisms are
compared with experiment in Fig.~\ref{fig:ad:he9}~(a) and
Fig.~\ref{fig:ad:he9}~(b). The experimental data present a sharp drop
with increasing angle, characteristic of an $L=0$
transition. Consequently, despite the very limited statistics, the
present data support the contention that the lowest lying state in
$^9$He is $1/2^+$.

The present work is also compared in Fig.~\ref{fig:he9:sum} with
experiments utilizing knock-out reactions to study $^9$He. For states
close to the neutron threshold results were obtained in terms of
scattering lengths: $a_s = -10$~fm \cite{Chen:2001qy}, $a_s \geq
-3$~fm \cite{alfalou:n7:conf} and $a_s = -3.17(66)$~fm
\cite{Johansson:2010fk}. Assuming that the low-lying structure
observed is a resonance, a corresponding energy $E_r$ is calculated
and shown in Fig.~\ref{fig:he9:sum}.  However, in this section we
prefer to compare scattering lengths and since the G.S. is close to
the neutron threshold we use the relation $E_r \approx
\frac{\hbar^2}{2\mu a_s^2}$~\cite{Chen:2001qy} (where $\mu$ is the
reduced mass for the neutron + $^8$He system) to obtain the
corresponding value $a_s \approx -12\pm 3$~fm for the scattering
length from this work. This scattering length is comparable to the
result of Chen~\emph{et al.}~\cite{Chen:2001qy} but is not compatible
with the weakly interacting $s$-wave strength found both by
Al~Falou~\emph{et al.}~\cite{alfalou:n7:conf} and Johansson~\emph{et
  al.}~\cite{Johansson:2010fk}. This may suggest, as noted by
Johansson~\emph{et al.}~\cite{Johansson:2010fk}, that the accumulation
of strength close to the neutron threshold observed in these two
experiments is inherent to the reaction and experimental conditions
and not the observation of a well defined $s$-wave G.S.

The weak binding energy approximation used to calculate the
theoretical angular distributions involves the use of a low binding
energy (here 0.0001~MeV) to enable the calculation of the form factor
in the usual way for unbound states while retaining the correct
excitation energy for the ``kinematical'' part of the calculation.
For $L=0$ states strong variations in the calculated absolute cross
section are observed as a function of the choice of the binding energy
and it is therefore impossible to extract meaningful spectroscopic
factors from the DWBA calculation in such cases. However, it is
possible to estimate a value from the single-particle width. Using the
prescriptions of Lane and Thomas \cite{rmatrix} we find $\Gamma_{sp}
\approx 2700$~keV for $E_r=180$~keV. Experimentally $\Gamma =
180\pm160$~keV, which corresponds to a spectroscopic factor smaller
than $\sim 0.13$. Our calculation may however be too crude and more
appropriate theoretical approaches are necessary to confirm this
estimation.

It is more difficult to deduce the nature of the first excited state
observed here at around 1.3~MeV above threshold from its angular
distribution. Within the experimental uncertainties both the $L=1$ and
$L=2$ calculations reproduce the data (Fig.~\ref{fig:ad:he9}~(c)).
Our angular distribution is compatible with the $J^\pi = 1/2^-$
spin-parity assigned in most of the previous studies
(Fig.~\ref{fig:he9:sum}). The small width measured here ($\Gamma
=130\pm170 $~keV) corroborates several previous results
\cite{oer95,Bohlen:1999qq,alfalou:phd,Johansson:2010fk,Rogachev:2003lq}. The
values of the corresponding spectroscopic factors (Tab.~\ref{tab:9He})
vary by a factor of up to 3 depending on the DWBA input parameters,
but it is worth noting that all of them are substantially smaller than
1 (of the order of 0.05 for $L=1$). This indicates that the first
excited state is of a strongly mixed nature in agreement with the
small observed width. From the analysis of this width, Barker
\cite{bar04} found spectroscopic factors $C^2S < 0.1$. Here, a
calculation using the Lane and Thomas prescription \cite{rmatrix}
gives a single-particle width of $2.4$~MeV for an $L=1$ resonance at
1.25~MeV.  From the observed width a spectroscopic factor of $C^2S
\approx 0.06$ is deduced, in agreement with that extracted from the
experimental angular distribution.

The excited state found here at around 3.5~MeV shows a smoothly
decreasing angular distribution of $L=2$ character
(Fig.~\ref{fig:ad:he9}~(d)). Due to the large uncertainties in its
energy ($\approx 800$~keV), this state could be compared to the
$5/2^+$ state found at around 4~MeV in
Refs.~\cite{Rogachev:2003lq,Golovkov:2007fu}. The small corresponding
spectroscopic factors suggest that this state is also strongly mixed.
However, we extracted a width of the order of 3~MeV.  Such a large
width for decay to the ground state of $^8$He would indicate a large
spectroscopic factor but the measured cross section is compatible with
an upper limit of $C^2S \lesssim 0.05$. We note that the cross section
would be even smaller for a particular state if several states are
present.  To render these observations compatible, considering that
this state is located above the $4n$-threshold, a strong decay branch
to the $^8$He $2^+$ state is necessary.

Finally, only a few theoretical calculations are available for $^9$He
and we concentrate here on recent results. It is interesting to note
that none of the recent calculations of the No Core Shell Model
\cite{Lisetskiy2008} or the Unified Shell Model Approach
\cite{Volya2005} predict parity inversion, both suggesting a $1/2^-$
ground state. Only Otsuka~\emph{et al.\/} \cite{Otsuka:PRL87}, within
a more standard shell-model framework, are able to account for the
inversion with the help of a strong attractive monopole interaction.

\section{Conclusions}
We have investigated the $^8$He(d,p) neutron stripping reaction in
inverse kinematics at 15.4$A$~MeV to explore the structure of the
unbound neutron-rich nucleus $^9$He.  Despite limited statistics the
angular distribution for the state observed very close to the
$^8\text{He}+\text{n}$ threshold ($180\pm85$~keV above) supports the
contention that the ground state spin-parity of $^9$He is $1/2^+$,
thus confirming the parity inversion.  Two excited states were
observed: the first lies approximately $1.3$~MeV above the neutron
threshold and is compatible with $J^\pi = 1/2^-$, exhibiting at the
same time a strong mixed nature. A second state at $\sim 3.5$~MeV
above threshold presents an $L=2$ angular distribution, suggesting a
$3/2^+$ or $5/2^+$ spin-parity assignment.

The present analysis suggests that in the excitation energy range up
to 5~MeV in $^9$He there is no state with a strong component of a
single neutron coupled to a core of $^8$He in its ground state. This
is a qualitative change from $^7$He, where the ground state has a
large spectroscopic factor for $^6\text{He}_\text{gs}$ coupled to a
single neutron.  The large width of the $3.5$~MeV state for decay to
the $2^+$ state in $^8$He is compatible with the assumption that this
state has a strong component of an excited
$^8\text{He}\left(2^+\right)$ core coupled to a single neutron.

\begin{acknowledgments}

We wish to thank the GANIL technical staff. NK and KR gratefully
acknowledge receipt of a COPIGAL grant. We would also like to thank
F.~Gulminelli and R.~Raabe for fruitful discussions and advice.

\end{acknowledgments}

\bibliographystyle{apsrev4-1}

\bibliography{he9-PRC}

\end{document}